\documentclass[aps,prl,a4paper,twocolumn,showpacs,floatfix,citeautoscript,superscriptaddress,footinbib]{revtex4-1}

\usepackage{graphicx}
\usepackage{amsfonts}
\usepackage{amsmath}
\usepackage{amssymb}
\usepackage{bm}

\usepackage{bbold}
\usepackage{mathtools}
\usepackage{dcolumn}
\usepackage{hyperref}
\usepackage[usenames]{color}
\usepackage{subcaption}

\usepackage{physics}
\usepackage{units}

\usepackage{ragged2e}
\DeclareCaptionJustification{justified}{\justifying}
\newcommand{\out}[1]{{}}

\newcommand{\fref}[1]{Fig.~\ref{#1}}

\begin{document}

\title{Dynamical control of Coulomb interactions and Hubbard bands in monolayer 1T-TaS$_2$}

\author{Niklas Notter}
\address{Institute of Theoretical and Computational Physics, TU Graz, Petersgasse 16, 8010 Graz, Austria}
\author{Markus Aichhorn}
\address{Institute of Theoretical and Computational Physics, TU Graz, Petersgasse 16, 8010 Graz, Austria}
\author{Anna Galler}
\email{anna.galler@tugraz.at}
\address{Institute of Theoretical and Computational Physics, TU Graz, Petersgasse 16, 8010 Graz, Austria}
\address{Max Planck Institute for the Structure and Dynamics of Matter, Center for Free Electron Laser Science, Luruper Chaussee 149, 22761 Hamburg, Germany}

\begin{abstract} 
Monolayer 1T-TaS$_2$ hosts a star-of-David charge-density wave (CDW) that stabilizes a low-temperature Mott-insulating state. Recent time-resolved spectroscopies indicate a coupling between the CDW amplitude mode and the electronic correlation strength, yet the role of the screened Coulomb interaction remains unclear.
Using the constrained random-phase approximation, we show that the CDW amplitude modifies the bare and screened on-site interactions, leading to sizable variations in the effective Hubbard $U$. Our combined density-functional and dynamical mean-field theory calculations reveal that the Hubbard bands shift in concert with the CDW amplitude, and that a reduced distortion drives a transition from a Mott insulator to a correlated metal.
These results demonstrate a direct link between lattice distortions and Coulomb interactions in transition-metal dichalcogenides, providing a microscopic mechanism for light-induced control of correlated phases in two-dimensional quantum materials.
\end{abstract}

\maketitle

Two-dimensional (2D) correlated materials exhibit a rich variety of quantum phases, including superconducting~\cite{saito2016}, charge- and spin-density wave~\cite{lin2020patterns} and quantum spin liquid states~\cite{ruan2021,Klanjsek2017}. The emergence of these phases is sensitive to external parameters such as pressure, strain, and static or dynamic electric and magnetic fields. Beyond this intrinsic tunability, 2D materials offer an additional degree of control through stacking and twisting, enabling emerging paradigms such as oxide electronics~\cite{sonsteby2020,keimer2013}, valleytronics~\cite{schaibley2016,cao2012valley}, and twistronics~\cite{andrei2020,Rubio2020twist,graph_twist}.

\begin{figure}[t!]
\includegraphics[width=\columnwidth]{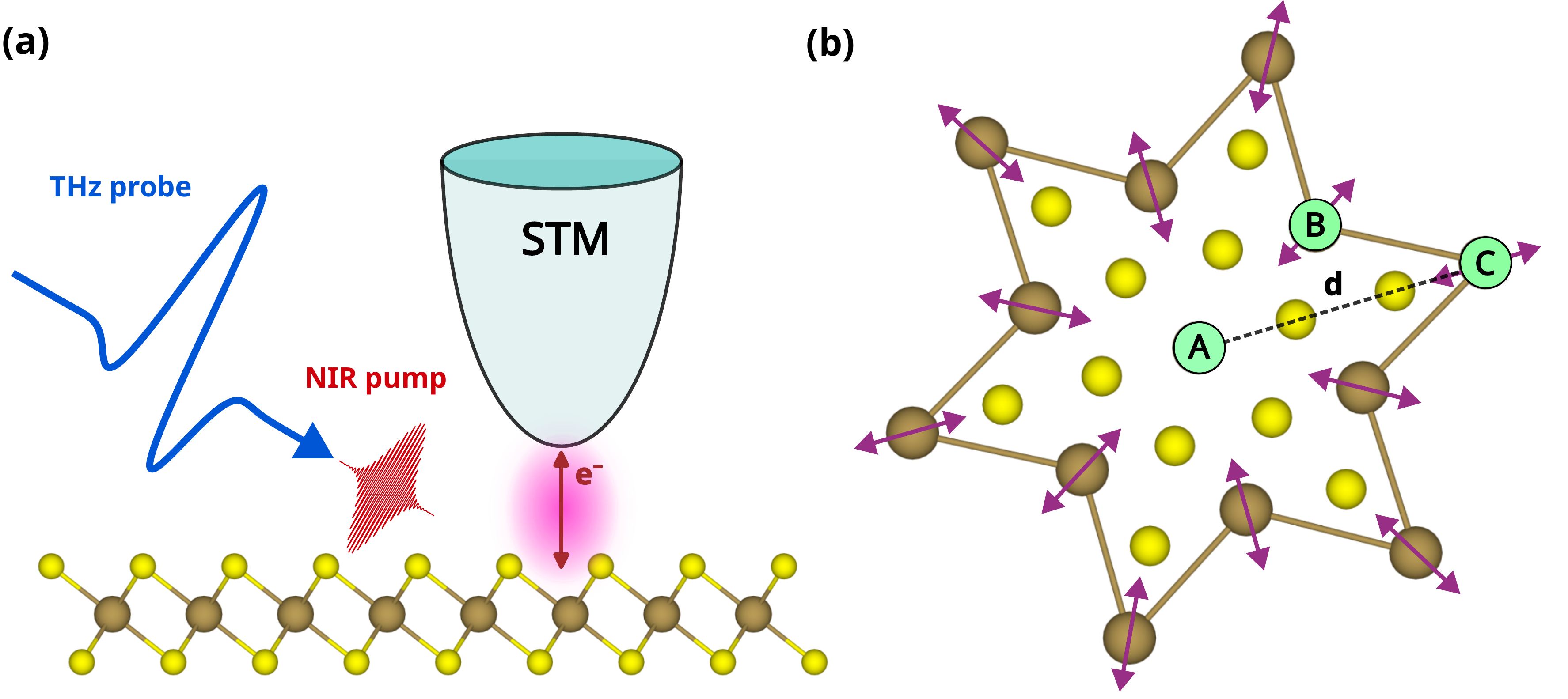}
\caption{\textbf{(a)} Illustration of a time-resolved STM experiment. A near-infrared (NIR) pulse indirectly excites the CDW amplitude mode in monolayer 1T-TaS$_2$, while a time-delayed THz probe pulse induces electron tunneling between the sample and STM tip to probe the local density of states. \textbf{(b)} Top view of the SoD supercell with Ta (brown) and S (yellow) atoms. $d$ denotes the distance between the central (site A) and the outer (site C) Ta atoms. Violet arrows indicate the CDW amplitude mode oscillation. } 
\label{fig:STM}
\end{figure}

Transition-metal dichalcogenides (TMDs) in the 1T polymorph~\cite{straub2024,petocchi2022,shin2021mott,crommie2020,nakata2016,marianetti2014,Perfetti2008,perfetti2003,Dardel1992,Claessen1990,Tosatti1979} are van-der-Waals materials that allow us to explore strong correlation effects in 2D. Among them, 1T-TaS$_2$ exhibits a Star-of-David (SoD) charge density wave (CDW) phase below \unit[180]{K}. The formation of this CDW phase, where clusters of 13 Ta atoms  move closer in a star-shaped arrangement (\fref{fig:STM}b), is accompanied by a transition to an insulating state. While bulk 1T-TaS$_2$ is believed to consist of  dimerized, band-insulating bilayers interspersed with AC-stacked monolayers~\cite{hua2025,wang2024,butler2020}, isolated monolayers  are well established as Mott insulators~\cite{wang2024,marianetti2014,perfetti2003}. The CDW distortion in a monolayer generates a very narrow, half-filled band at the Fermi level. This molecular orbital, with meV bandwidth and predominantly Ta $d_{3z^2-r^2}$ character, is highly susceptible to electronic correlations.
To first approximation, monolayer 1T-TaS$_2$ can be regarded as a prototypical realization of the Hubbard model on a triangular lattice, a system that recently attracted  interest due to the interplay of strong correlations, geometric frustration, and unconventional ordering~\cite{vandelli2024,wietek2021,Weitering2020_SC,Hansmann2019_SC,He2018}.
 
The CDW distortion clearly plays a central role in stabilizing the correlated state, suggesting that it could serve as a control knob to manipulate the electronic properties. Initial experiments indicated that external strain, which modifies the amplitude of the SoD CDW, can influence the Mott transition~\cite{arxiv_dark_stars}. More recent studies using terahertz (THz) scanning tunneling microscopy (STM)~\cite{lopez2025} and time- and angle-resolved photoemission spectroscopy (tr-ARPES)~\cite{bovensiepen2025,maklar2023,hellmann2012,perfetti2006}, which excite a coherent \unit[2.4]{THz} CDW amplitude mode~\cite{demsar2002} in 1T-TaS$_2$, point toward the possibility of dynamically controlling the correlation strength. However, despite recent density-functional theory (DFT) and non-equilibrium dynamical mean-field theory (DMFT) calculations~\cite{bovensiepen2025}, the role of the screened Coulomb interaction in the photoinduced dynamics remains insufficiently understood. This limitation primarily stems from the difficulty of quantitatively evaluating the screened interaction within the large SoD supercell.

\begin{figure*}[t!]
\begin{minipage}{18. cm}
\includegraphics[width=\columnwidth]{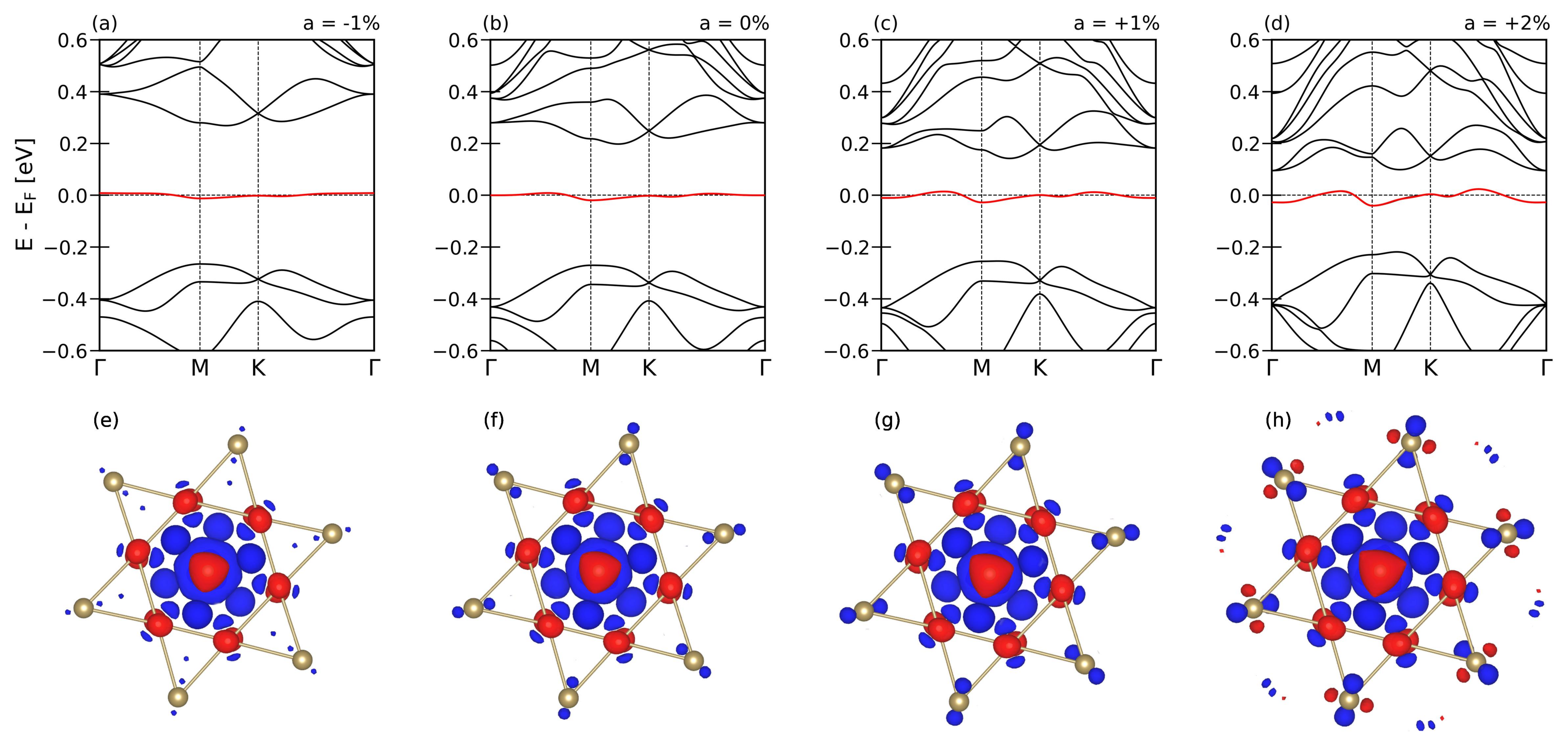}
\caption{\textbf{DFT band structure and Wannier projection for varying CDW distortion.} \textbf{(a–d)} Electronic band structure of monolayer 1T-TaS$_2$ for CDW amplitudes $a=-1\%$ to $a=+2\%$ around equilibrium, computed with PBE~\cite{Perdew1996} in VASP~\cite{Kresse1996b}. The half-filled band at the Fermi level (red) is projected onto a maximally localized Wannier function~\cite{Pizzi2020}.
\textbf{(e–h)} Wannier isosurfaces centered on the central Ta atom with smaller weight on surrounding atoms. Red/blue denote positive/negative values. The orbital becomes increasingly delocalized with larger CDW amplitude.} 
\label{fig:wann}
\end{minipage}
\end{figure*}

Using first-principles calculations within the constrained random-phase approximation (cRPA)~\cite{aryasetiawan2004,Kotani2000,springer1998}, we show that the screened Coulomb interaction in 1T-TaS$_2$ is highly sensitive to the amplitude of the SoD CDW. Increasing the CDW amplitude reduces the localization of the Ta $d_{3z^2-r^2}$ molecular orbital, lowering the bare Coulomb interaction, while enhanced electronic screening further decreases the equilibrium on-site interaction of 
$U$=\unit[0.4]{eV} by up to \unit[0.1]{eV} for a 1\% amplitude change (see below). Consequently, the Hubbard bands shift and the Mott gap shrinks, and strikingly, a  3\% increase in the CDW amplitude drives a transition from a Mott insulator to a correlated metal. These results establish a direct, quantitative link between structural distortions and electronic correlations in TMDs, providing a route toward optical control of correlated phases and guiding the interpretation of ultrafast spectroscopic experiments in 2D CDW materials.

\section{Results}
We start by describing the idealized pump–probe setup underlying our simulations, illustrated in \fref{fig:STM}. The configuration resembles a THz-lightwave–driven STM experiment~\cite{lopez2025}, where a near-infrared (NIR) laser pulse excites coherent phonon dynamics in a monolayer of 1T-TaS$_2$ through photoinduced charge redistribution. In particular, the pulse drives the \unit[2.4]{THz} CDW amplitude mode~\cite{demsar2002}, schematically depicted in \fref{fig:STM}b. This breathing phonon mode corresponds to oscillations of the outer Ta atoms (sites B and C) around their equilibrium positions within the SoD CDW structure. Throughout this work, we express the oscillation amplitude $a$ as a percentage of the distance $d$ between the central (site A) and outer (site C) Ta atoms in the SoD cluster, i.e. $a=(d-d_\mathrm{eq})/d_\mathrm{eq}$, where $d_\mathrm{eq}$ is the distance in the equilibrium CDW distortion. A positive $a$ thus indicates an expanded star, corresponding to a configuration closer to the undistorted phase (which corresponds to $a=6.5\%$), while a negative $a$ denotes a contracted star with enhanced CDW distortion.  The resulting lattice vibration modulates the electronic states---a key effect we aim to quantify in this study. The time-dependent electronic spectral function can be probed by a THz pulse, which induces electron tunneling between the STM tip and the sample, as depicted in \fref{fig:STM}a. By varying the time delay between the NIR and THz pulses, one can track the evolution of the electronic structure over an entire CDW amplitude-mode oscillation.

In our simulations, we model the coherent amplitude-mode oscillation of the SoD CDW in 1T-TaS$_2$ using a frozen-phonon approach that decouples electronic and ionic degrees of freedom. The amplitude $a$  is modulated, and the corresponding electronic structure is evaluated separately for each frozen ionic configuration. This treatment assumes that the electronic response, which occurs on femtosecond timescales, rapidly adapts to each lattice configuration. While not accounting for electron-phonon or phonon-phonon interactions, we expect this approach captures the essential changes of the electronic structure over the  THz CDW amplitude mode oscillations, where one period lasts approximately \unit[0.4]{ps}.

\begin{figure*}[t!]
\begin{minipage}{18. cm}
\includegraphics[width=\columnwidth]{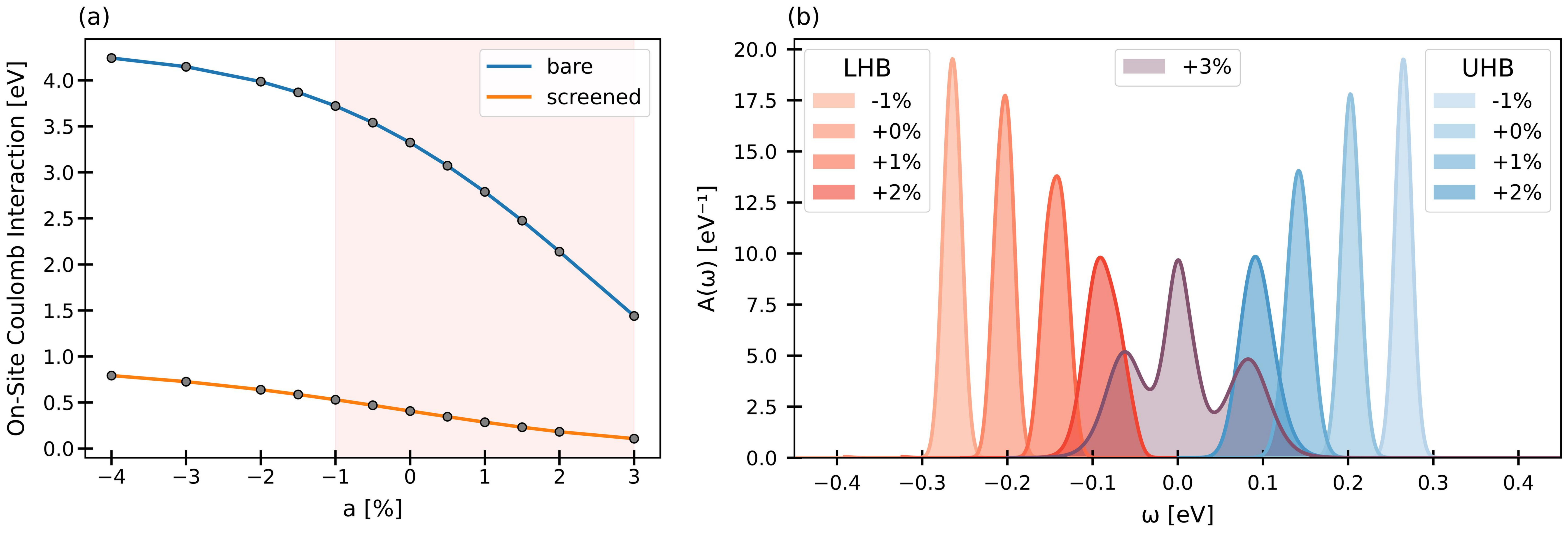}
\caption{\textbf{Tuning the screened Coulomb interaction and Mott gap via CDW amplitude modulation.} \textbf{(a)} Bare and screened on-site Coulomb interaction as a function of CDW amplitude $a$. The shaded region highlights the parameter range for which the electronic spectral functions $A(\omega)$ are shown in
  \textbf{(b)}. Increasing the CDW amplitude from -1\% to +2\% shifts the lower (LHB) and upper Hubbard bands (UHB) and reduces the Mott gap. At  $a=3\%$, the system undergoes a transition to a correlated metallic state. In (b) the chemical potential was set to the center of the gap.} 
\label{fig:coulomb}
\end{minipage}
\end{figure*}

In \fref{fig:wann}, we present the DFT band structure of monolayer 1T-TaS$_2$, computed for different amplitudes $a$ of the CDW distortion, ranging from $a=-1\%$ to $a=+2\%$. The reference case $a=0\%$ (\fref{fig:wann}b) corresponds to the equilibrium SoD CDW. A prominent feature in all cases is the emergence of a very narrow, half-filled band at the Fermi level (highlighted in red) with a bandwidth of just a few tens of meV. Using \textsc{wannier90}~\cite{Pizzi2020,Mostofi2008,Kresse1996b}, we project this band onto a maximally localized Wannier orbital, which turns out to be a molecular-like state with predominant Ta $d_{3z^2-r^2}$ character. As shown in \fref{fig:wann}e–h, the Wannier orbital is centered on the central Ta atom (site A) but extends significantly onto the surrounding Ta atoms of the SoD cluster. Upon increasing the CDW amplitude from $a=-1\%$ (\fref{fig:wann}e) to $a=+2\%$ (\fref{fig:wann}h), the Wannier orbital becomes progressively more delocalized, with its spread increasing from \unit[24]{\AA$^2$} to \unit[95]{\AA$^2$}. This enhanced delocalization directly correlates with an increase in bandwidth of the half-filled band, which grows from \unit[20]{meV} (\fref{fig:wann}a) to  \unit[70]{meV} (\fref{fig:wann}d).

The narrow, half-filled band is highly susceptible to electronic correlations, making it essential to evaluate the screened Coulomb interaction within this correlated subspace. In the random-phase approximation (RPA), the fully screened interaction is given by $W=(1-V\chi_0)^{-1}V$, with the bare interaction $V$ and the independent-particle polarization $\chi_0$. In constrained RPA (cRPA)~\cite{aryasetiawan2004,Kotani2000,springer1998}, $\chi_0$ is  decomposed into contributions from the correlated target subspace ($\chi_0^t$) and the remainder $\chi_0^r$, yielding the partially screened interaction~\cite{nomura2012,kutepov2010} 
\begin{equation}
U = W(1+\chi_0^tW)^{-1},
\end{equation}
where $\chi_0^t$ is the static, independent-particle polarization in the target space 
\begin{align}
\chi_0^t(\textbf{r},\textbf{r}^\prime) = &2\sum_{nn^\prime\in t}\sum_{\textbf{q}\textbf{k}}\frac{f_{n^\prime\textbf{k}+\textbf{q}}-f_{n\textbf{k}}}{\epsilon_{n^\prime\textbf{k}+\textbf{q}}-\epsilon_{n\textbf{k}}}\psi^*_{n\textbf{k}}(\textbf{r}) \\\nonumber
&\times\psi_{n^\prime\textbf{k}+\textbf{q}}(\textbf{r})\psi^*_{n^\prime\textbf{k}+\textbf{q}}(\textbf{r}^\prime)\psi_{n\textbf{k}}(\textbf{r}^\prime),
\end{align}
with $\{\psi_{n\textbf{k}},\epsilon_{n\textbf{k}}\}$ denoting single-particle wave functions and energies, and 
$f$ the Fermi functions.
We employ the projector cRPA method as implemented in VASP, defining the target space via the Wannier projection. Calculations are performed in the SoD supercell, using 1024 bands and 8×8×1 
$\textbf{k}$-points to achieve convergence (further computational details are provided in the Supplementary Information).

\fref{fig:coulomb} presents the central result of this work. Panel (a) shows the bare and screened on-site Coulomb interactions in the molecular Ta $d_{3z^2-r^2}$ Wannier orbital for CDW amplitudes $a$ ranging from -4\% to +3\% around the equilibrium value ($a=0\%$). The bare interaction $V$ decreases markedly from \unit[4.2]{eV} to \unit[1.4]{eV} as $a$ increases from –4\% to +3\%, reflecting the progressive delocalization of the Wannier orbital: as the orbital spreads out, the Coulomb repulsion is reduced. In parallel, electronic screening becomes stronger due to global changes in the band structure. This trend is evident in the DFT band structures shown in \fref{fig:wann}a-d, where larger CDW amplitudes bring the valence-band maximum and conduction-band minimum closer together, enhancing screening in the narrow correlated band. The resulting screened interaction $U$ is plotted in orange in \fref{fig:coulomb}a. At the equilibrium distortion ($a=0\%$), we obtain $U = \unit[0.4]{eV}$. Around this point, $U$ varies approximately linearly, changing by about \unit[0.1]{eV} for $a=\pm1\%$.

Next, we employ the screened Coulomb interaction $U$ obtained from cRPA to compute the electronic spectral function of monolayer 1T-TaS$_2$ as a function of the CDW distortion. For each CDW distortion with amplitude $a$,  we derive a single-band Hubbard model,    
\begin{equation}
\hat{H}_a = \sum_{ij,\sigma}H_a^{(\textbf{R}_i-\textbf{R}_j)}\hat{c}^\dagger_{i\sigma} \hat{c}_{j\sigma}+U_a\sum_{i}\hat{n}_{i\uparrow} \hat{n}_{i\downarrow},
\end{equation}
where $H_a^{(\textbf{R}_i-\textbf{R}_j)}$ represents the single-particle Hamiltonian in the Wannier basis,  and $U_a$ denotes the screened on-site interaction in the correlated Ta $d_{3z^2-r^2}$ orbital. The operators $\hat{c}^\dagger_{i\sigma}$ and $\hat{c}_{i\sigma}$  create and annihilate an electron with spin $\sigma$ at lattice site $\textbf{R}_i$, respectively, and $\hat{n}_{i\sigma}=\hat{c}^\dagger_{i\sigma}\hat{c}_{i\sigma}$. 
The resulting model is solved within DMFT using the continuous-time quantum Monte Carlo (CT-HYB) impurity solver implemented in w2dynamics~\cite{wallerberger2019}.
All computations are performed at an inverse temperature $\beta=\unit[1200]{eV^{-1}}$, corresponding to approximately \unit[10]{K}. Analytical continuation to real frequencies is carried out using the maximum entropy method as implemented in the ana\_cont package~\cite{kaufmann2023ana}. Further technical details of the DFT+DMFT calculations are provided in the Supplementary Information.

In equilibrium ($a = 0\%$), the electronic spectral function $A(\omega)$ of monolayer 1T-TaS$_2$ is  split into lower (LHB) and upper Hubbard bands (UHB), as shown in \fref{fig:coulomb}b. These bands are separated by a Mott gap of approximately \unit[0.4]{eV}. Upon excitation of the CDW amplitude mode, i.e., when $a$ is varied, the Hubbard bands exhibit a pronounced oscillatory behavior: for negative $a$ (enhanced distortion), the bands move further apart, while for positive $a$ (reduced distortion), the Mott gap narrows. The evolution of the Mott gap closely follows the variation of the screened interaction $U_a$ obtained from cRPA.

At $a=+3\%$, a transition to a correlated metallic state occurs, as evidenced by the emergence of a pronounced quasiparticle peak at the Fermi level, coexisting with reminiscent LHB and UHB features. These results demonstrate that the screened Coulomb interaction, Mott physics, and metal-to–insulator transition in monolayer 1T-TaS$_2$ can be effectively tuned by the CDW amplitude. Furthermore, since the Hubbard band positions oscillate in phase with the CDW amplitude mode, this mechanism opens a route to dynamically control electronic correlations through lightwave-driven lattice vibrations.

\begin{figure*}[t!]
\begin{minipage}{18. cm}
\includegraphics[width=\columnwidth]{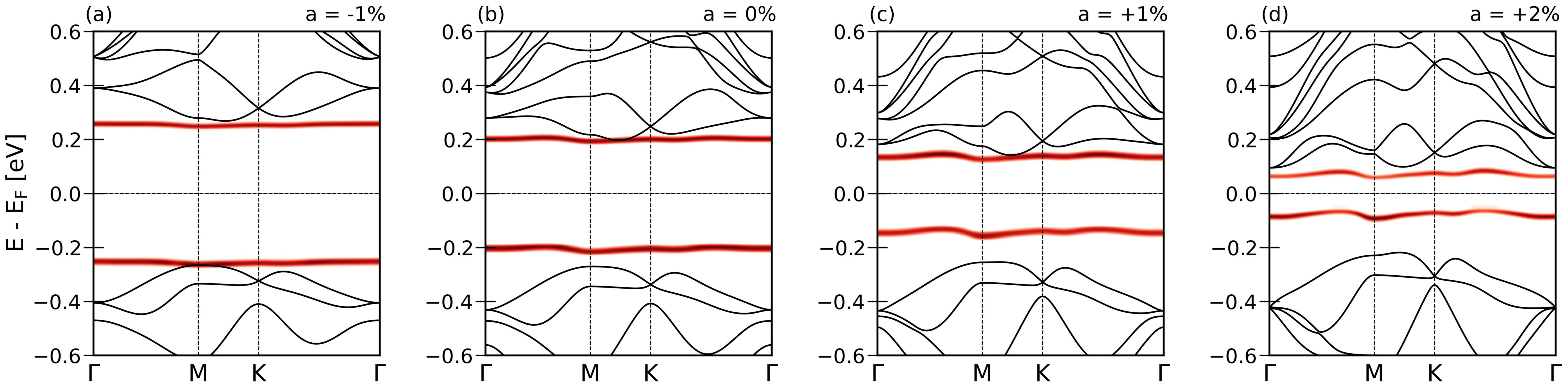}
\caption{\textbf{Momentum-resolved spectral function as a function of CDW amplitude.} The electronic spectral function of the Ta $d_{3z^2-r^2}$ orbital (red) obtained from DFT+DMFT is shown together with the DFT band structure (black). As the CDW amplitude $a$ is varied from \textbf{(a)} -1\% to \textbf{(d)} +2\%, the Mott gap---defined as the spectral gap between the occupied LHB and the unoccupied UHB---gradually decreases from 0.5 to 0.1 eV.    } 
\label{fig:dmft}
\end{minipage}
\end{figure*}

Finally, \fref{fig:dmft} presents the $\textbf{k}$-resolved electronic spectral function (in red), obtained from DMFT and overlaid with the DFT band structure. The alignment between the correlated Ta $d_{3z^2-r^2}$ band and the remaining states was set using a fully localized limit double-counting correction~\cite{anisimov1993}, $E_{\mathrm{DC}} = U_a / 2$. The panels (a-d) show spectra for CDW amplitudes from $a = -1\%$ to $a = +2\%$. As in the $\textbf{k}$-integrated spectra of \fref{fig:coulomb}b, the Ta $d_{3z^2-r^2}$ orbital splits into LHB and UHB, and the Mott gap narrows  as the CDW amplitude increases. The predicted $\textbf{k}$-resolved spectral signatures can be probed by tr-ARPES experiments, where coherent excitation of the CDW amplitude mode can dynamically tune the Mott state.

\section{Discussion}
Discovering effective ways to tune correlated phases in quantum materials remains a central challenge in condensed matter physics, with significant implications for future technologies.
Here, we demonstrate a direct and quantitative link between structural distortions and electronic correlations in monolayer 1T-TaS$_2$. Using a combination of ab-initio and many-body simulations, we show that the CDW amplitude $a$ acts as an efficient tuning parameter for the screened Coulomb interaction and thus for the overall correlation strength. A variation of $a=±1\%$ leads to substantial shifts of the Hubbard bands and a change of the Mott gap up to \unit[0.1]{eV}, highlighting a strong coupling between lattice and electronic degrees of freedom.
These predicted changes exceed those observed experimentally~\cite{lopez2025,maklar2023,perfetti2006}, which we attribute to (i) enhanced screening effects in bulk samples, where adjacent layers reduce the effective Coulomb interaction, and (ii) smaller amplitude-mode oscillations achieved in pump–probe experiments. Importantly, our simulations provide a quantitative link between Hubbard-band shifts and CDW amplitude, offering experimentalists a direct route to extract CDW amplitudes from spectroscopic data.

Our first-principles evaluation of the screened Coulomb interaction $U$ provides  essential microscopic input for interpreting time-resolved spectroscopies of 1T-TaS$_2$ and related materials. It establishes a direct pathway for manipulating Mott insulating states in 2D CDW systems via coherent lattice dynamics.
While our equilibrium simulations isolate the impact of structural modulations and do not capture the ultrafast nonequilibrium regime immediately following photoexcitation, they describe the relevant electronic states on longer (\unit[100]{fs}--\unit[1]{ps}) timescales, where electrons have thermalized and respond adiabatically to the lattice’s \unit[2.4]{THz} amplitude mode oscillation. Future extensions including coupled electron-phonon dynamics and nonlocal Coulomb interactions may further refine this picture.
Overall, our work uncovers how CDW lattice distortions can govern electronic correlations, offering both a conceptual framework and a practical guide for engineering correlated phases in low-dimensional quantum materials.

 \paragraph{Acknowledgments} We thank Melanie M\"uller, Jan M. Tomczak, and Chia-Nan Yeh for helpful discussions. This work was financially supported by the Austrian Science Fund (FWF) grant 10.55776/V988. The computational results have been achieved using the Austrian Scientific Computing (ASC) infrastructure.

%

\end{document}